\documentclass[english,aps,prd,preprint]{revtex4}
\usepackage[latin1]{inputenc}
\usepackage{graphicx}

\makeatletter

\providecommand{\tabularnewline}{\\}

\usepackage{babel}
\makeatother
\begin{document}

\title{Heavy lepton production at Linac$\otimes$LHC}

\author{A. T. Alan%
\footnote{E-mail: alan\_a@ibu.edu.tr%
} and A. T. Tasci%
\footnote{E-mail: tasci\_a@ibu.edu.tr%
} }

\affiliation{Abant Izzet Baysal University, Department of Physics,
14280, Bolu, Turkey}

\author{O. Çakir%
\footnote{E-mail: ocakir@science.ankara.edu.tr%
}}

\affiliation{Ankara University, Department of Physics, 06100,
Tandogan, Ankara, Turkey}

\begin{abstract}
We investigate the production, signatures and backgrounds of new
heavy leptons via string inspired $E_{6}$ model at the proposed
Linac$\otimes$LHC. Assuming maximal mixing, the production rate is
found to be 2000 events per year for masses up to 3 TeV.
\end{abstract}
\maketitle

\section{introduction}

Although the Standard Model (SM) is very successful in explaining
the physics of many phenomena under a few hundred GeV, it is not a
complete description of the physics at higher energies. The most
fundamental problems with SM are the mass hierarchy and the number
of fermion generations. There are many models extending the SM to
overcome these problems. Almost all of these new models include
new fermions in addition to the ordinary ones. We investigate the
possibility of the single production of new heavy leptons
suggested by string inspired $E_{6}$ model in $ep$ collisions.
There are many analysis of the heavy lepton production at future
linear colliders \cite{key-1,key-2,key-3}, at hadron colliders
\cite{key-4,key-5,key-6} and also at $ep$ collider HERA
\cite{key-7}. For the searches of new physics beyond the SM, the
linac-ring type $ep$ colliders have as much potential as lepton
colliders \cite{key-8}.

\section{production of heavy leptons}

The model that we use in the single production of a new heavy
lepton is the string inspired $E_{6}$ model \cite{key-9,key-10}.
We therefore assume the new heavy lepton interactions in the
following flavor changing neutral current (FCNC) Lagrangian:
\begin{eqnarray*}
\mathcal{L}_{\mathtt{nc}}=g_{z}\sin\theta_{mix}\psi_{L}\gamma^{\mu}(1+\gamma_{5})\psi_{e}Z^{\mu}+h.c.
\end{eqnarray*}
and similar terms for the other leptonic families. Here
$\theta_{mix}$ are the mixing angles between right handed
components of the ordinary and new heavy charged leptons.
Throughout this paper we will suppose an upper limit
$\sin\theta_{mix}<0.1$ coming from the high precision measurements
of the $Z$ properties at LEP/SLC \cite{key-11}. We use the
parameter $b_{lLZ}$ for $\sin\theta_{mix}$ to denote the mixing in
the vertex $l-L-Z$ explicitly. In $E_{6}$, the parton level
process $eq\rightarrow Lq$, responsible for the heavy lepton
production in $ep$ collision occurs via FCNC $Z$ exchange in the
$t$-channel.

The differential cross section for the subprocess $eq\rightarrow
Lq$ in the framework of $E_{6}$ model is given by

\begin{eqnarray*}
\frac{d\hat{\sigma}}{d\hat{t}} & = & \frac{\pi\alpha^2 b_{lLZ}^{2}}{\sin^4\theta_W \cos^4 \theta_W\hat{s}^2[(\hat{t}-M_Z^2)^2+M_Z^2\Gamma_Z^2]}\left[(a_q+v_q)^2\hat{t}^2\right.\\
 &  & \left.+(a_q+v_q)^2(2\hat{s}-m_L^2)\hat{t}+2\hat{s}(a_q^2+v_q^2)(\hat{s}-m_L^2)\right],\end{eqnarray*}
where $\theta_W$ is the weak angle, $\alpha$ is the fine structure
constant, $\hat{s}$ and $\hat{t}$ are the Mandelstam variables and
$\hat{s}=xs$ is the square of the center of mass energy for the
subprocess while $x$ is the momentum fraction of the parton inside
the proton. The total cross section can be obtained by folding the
subprocess cross section $\hat{\sigma}$ over the parton
distribution functions as

\[
\sigma(ep\rightarrow
LqX)=\int_{x_{min}}^{1}dxf_{q}(x,Q^{2})\int_{t_{min}}^{t_{max}}\frac{d\hat{\sigma}}{d\hat{t}}d\hat{t}\]
where $x_{min}=m_{L}^{2}/s$, $\hat{t}_{min}=-(\hat{s}-m_{L}^{2})$
and $\hat{t}_{max}=0$. These relations are obtained for the
massless lepton and quark case. We give the production cross
sections for the signal as function of the heavy lepton mass,
$m_{L}$, in Fig. 1 for different values of $b_{lLZ}$. In Figure 2,
we display the invariant mass distribution of the background
process $ep\rightarrow qZeX$ as function of invariant mass of $Ze$
system at future lepton-hadron collider Linac$\otimes$LHC with the
main parameters $\sqrt{s}=5.3$ TeV and
$\mathcal{L=}10^{33}~cm^{-2}s^{-1}$ \cite{key-12}. We have used
the COMPHEP package \cite{key-13} to calculate the cross sections,
decay widths and branching ratios. For the parton distribution
functions $f_{q}(x,Q^{2})$ we have used MRS \cite{key-14} with the
factorization scale $Q^{2}=m_{L}^{2}$.

\begin{figure}
\includegraphics[%
  scale=0.5,
  angle=-90]{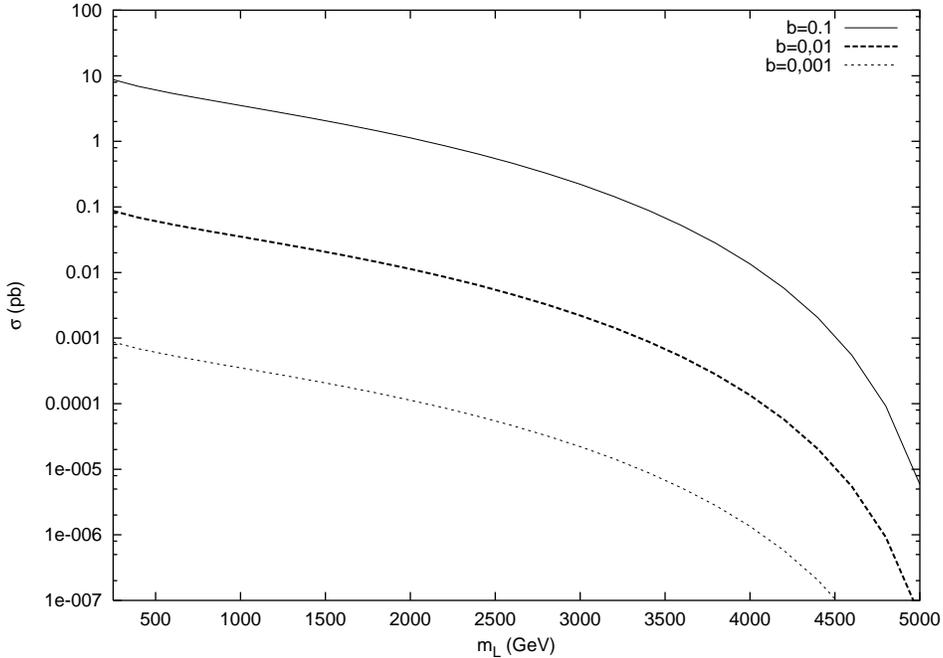}

\caption{Total production cross sections as functions of the heavy
lepton masses ($m_L$) at $\sqrt{s}=5.3$ TeV for different $l-L-Z$
couplings $b$. }
\end{figure}

\begin{figure}
\includegraphics{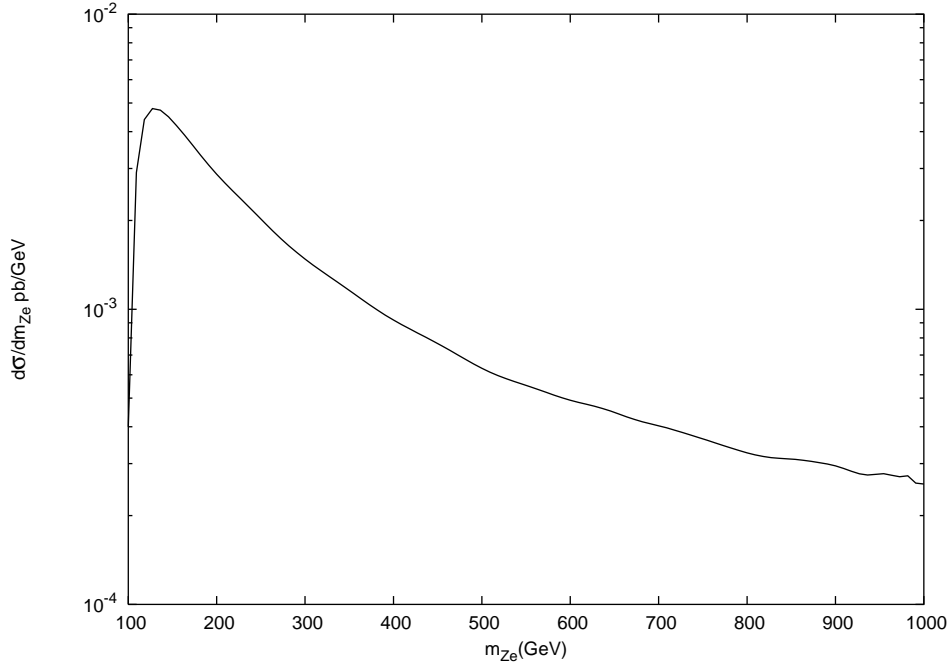}

\caption{The invariant mass distribution of the $Ze$ system for the background
process $ep\rightarrow qZeX$. }
\end{figure}

The heavy lepton production cross sections ($\sigma\times BR$) and
the number of signal events depending on the mass $m_{L}$ are
shown in Table 1. For decreasing values of the $l-L-Z$ couplings,
$b$, the production cross section and therefore the number of
events decreases.

\begin{table}

\caption{Cross sections depending on the heavy lepton mass $m_{L}$ for $b=0.1$.
The branching ratios $BR_{1}$ and $BR_{2}$ denote $BR(L\rightarrow Ze)$
and $BR(Z\rightarrow e^{+}e^{-},\mu^{+}\mu^{-})$, respectively. The
total decay width of the heavy lepton is given in the last column.}

\begin{tabular}{|c|c|c|c|c|}
\hline
$m_{L}$(GeV)&
$\sigma$(pb)&
$\sigma\times BR_{1}$(pb)&
$\sigma\times BR_{1}\times BR_{2}$(pb)&
$\Gamma$(GeV)\tabularnewline
\hline
\hline
200&
9.423&
3.109&
0.105&
$0.59$\tabularnewline
\hline
400&
6.873&
2.267&
0.076&
$5.18$\tabularnewline
\hline
600&
5.384&
1.776&
0.059&
$17.53$\tabularnewline
\hline
800&
4.333&
1.429&
0.048&
$41.53$\tabularnewline
\hline
1000&
3.519&
1.161&
0.039&
$81.05$\tabularnewline
\hline
2000&
1.129&
0.373&
0.012&
$647.16$\tabularnewline
\hline
3000&
0.221&
0.073&
0.002&
$2142.88$\tabularnewline
\hline
\end{tabular}
\end{table}

After their production, heavy leptons will decay via the neutral
current process $L\rightarrow lZ$, where $l$ is a light lepton (e,
$\mu$, $\tau$). The branching ratio for these processes would be
around $33 \%$ for each channel.

The backgrounds for the signal process $ep\rightarrow LqX$ with
the subsequent decays $L\rightarrow Z\mu$ or $L\rightarrow Z\tau$
and $Z\rightarrow e^+e^-$ are expected to be at very low rate. By
applying appropriate cuts to the final state particles this type
of backgrounds can be kept at very low levels. Still we may need
at least 10 signal events in the final state after all cuts.
Therefore, the $ep$ collider Linac$\otimes$LHC can probe heavy
lepton masses up to about 3 TeV as can be deduced from Table I.
For a heavy lepton with a mass of 200 GeV we expect $10^{3}$
signal events for the coupling value of $b=0.1$.

We applied an initial cut on the electron and jet transverse
momentum $p_{T}^{e,q}>10$ GeV for the signal and background
analysis. These cuts reduce the background by about 20\%. The
total background cross section is ($\sigma\times BR$)=0.055 pb
after the cuts. This improves the statistical significance
$S/\sqrt{B}$, where $S$ and $B$ denote the total signal and
background events, respectively. The calculated $3\sigma$ and
$5\sigma$ discovery contours for heavy lepton masses and
couplings, are displayed in Fig. 3.

We also have performed the same calculations for the THERA
($\sqrt{s}=1$ TeV, $\mathcal{L}=40~pb^{-1}$) collider
\cite{key-15}. Unfortunately, we have seen that at this collider
it is not possible to observe a heavy lepton with mass greater
than 200 GeV. Taking the $l-L-Z$ coupling value of 0.1 for a heavy
lepton with a mass of 200 GeV, the production rate is 100 events
per year.

\begin{figure}
\includegraphics{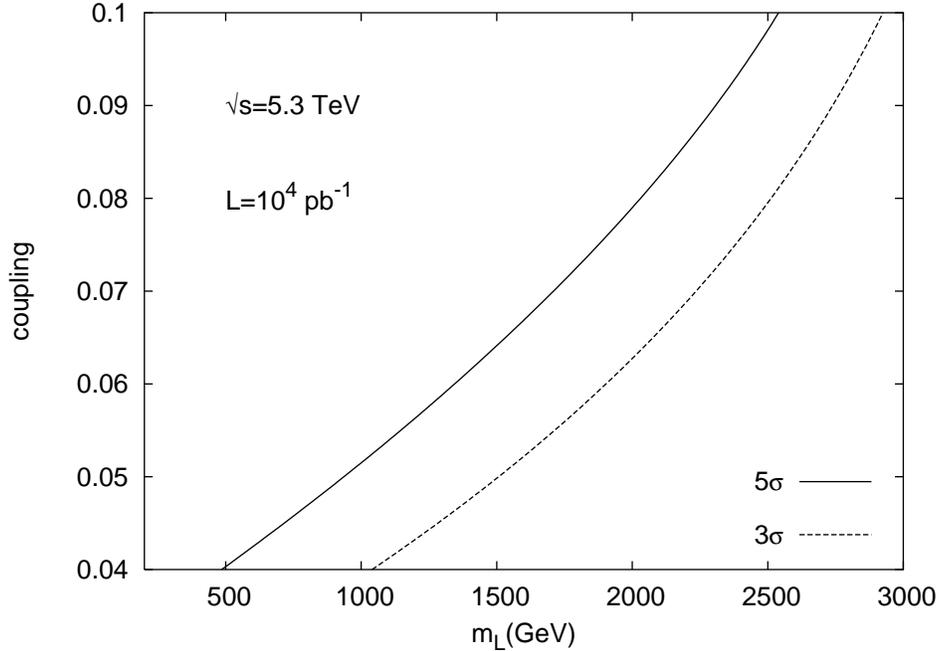}

\caption{Attainable mass limits depending on the coupling $b$ for heavy leptons
at Linac$\otimes$LHC.}
\end{figure}

\section{conclusions}

This work shows that some of future high energy lepton-hadron
colliders can test the existence of heavy leptons.
Linac$\otimes$LHC has very promising discovery potential for heavy
leptons with masses up to 3 TeV at $3\sigma$ significance, that
is, it offers the opportunity of the manifestations of new physics
beyond the SM, while at THERA it does not seem to be likely to
achieve masses greater than 200 GeV.

\begin{acknowledgments}
This work was partially supported by Abant Izzet Baysal University
Research fund.
\end{acknowledgments}

\end{document}